\documentclass{iaus}
\usepackage{graphicx}
\usepackage[breaklinks,dvipdfm]{hyperref}

\newcommand{\cobold}{{\sf CO$^5$BOLD}}

\def\aj{\textit{AJ}}%
\def\araa{\textit{ARAA}}%
\def\apj{\textit{ApJ}}%
\def\apjl{\textit{ApJ}}%
\def\apss{\textit{Ap\&SS}}%
\def\aap{\textit{A\&A}}%
\def\mnras{\textit{MNRAS}}%
\def\memsai{\textit{MemSAI}}
\def\pasj{\textit{PASJ}}%
\def\nat{\textit{Nature}}%
\title[The first galactic stars] 
{The first Galactic stars and chemical enrichment in the halo}

\author[Piercarlo Bonifacio]   
{Piercarlo Bonifacio$^{1,2,3}$}

\affiliation{$^1$ CIFIST Marie Curie Excellence Team \\[\affilskip]
$^2$ GEPI, Observatoire de Paris, CNRS, Universit\'e Paris Diderot; Place
Jules Janssen 92190
Meudon, France
\\email: {\tt Piercarlo.Bonifacio@obspm.fr}\\[\affilskip]
$^3$Istituto Nazionale di Astrofisica --
Osservatorio Astronomico di Trieste, Via Tiepolo 11,
I-34131 Trieste, Italy
}

\pubyear{2009}
\volume{265}  
\pagerange{81--90}
\jname{Chemical Abundances in the Universe: Connecting First Stars to Planets}
\editors{K. Cunha, M. Spite \& B. Barbuy, eds.}
\begin{document}

\maketitle

\begin{abstract}
The cosmic microwave background and the cosmic expansion can
be interpreted as evidence that the Universe underwent an
extremely hot and dense phase about 14 Gyr ago. The nucleosynthesis
computations tell us that the Universe emerged from this state
with a very simple chemical composition: H, $^2$H, $^3$He, 
$^4$He,  and traces
of $^7$Li. All other nuclei where synthesised at later times.
Our stellar evolution models tell us that, if a low-mass star
with this composition had been created (a ``zero-metal'' star) at 
that time,
it would still be shining on the Main Sequence today.
Over the last 40 years there have been many efforts to detect 
such primordial stars but none has so-far been found. The lowest
metallicity stars known have a metal content, $Z$, which is of the
order of $10^{-4}Z_\odot$. These are also the lowest metallicity
objects known in the Universe. This seems to support the 
theories of star formation which 
predict that only high mass stars could form with a primordial
composition and require a minimum metallicity to  
allow the formation of low-mass stars. 
Yet, since absence of evidence is not evidence of absence, we cannot
exclude the existence of such low-mass zero-metal stars, at present.
If we have not found the first Galactic stars, as a by product of our
searches we have found their direct descendants, stars of extremely low
metallicity  ($Z\le 10^{-3}Z_\odot$). The chemical composition 
of such stars contains indirect information on the nature of the
stars responsible for the nucleosynthesis of the metals.
Such a fossil record allows us a glimpse of the Galaxy at
a look-back time equivalent to redshift $z=10$, or larger.
The last ten years have been full of exciting discoveries
in this field, which I will try to review in this contribution.
\end{abstract}

\firstsection 
\section{Introduction}

The quest for the First Stars and their immediate
descendants has been  a field of very active research,
both in the high redshift and in the local Universe.
In this review I will only deal with advances in the local
Universe, mainly focusing on literature which appeared
in the last four years. I refer the reader to 
the reviews of \cite[Beers \& Christlieb (2005)]{BC05} and 
\cite[Bonifacio (2007)]{Bonifacio07} for the earlier
literature. I also largely omit the results on neutron-capture
elements, since this is covered by another review
in this Symposium (\cite[Sneden et al. 2009]{Sneden09}) and 
I refer to the review \cite[Sneden et al.(2008)]{Sneden08}  
for older literature.
I will touch only briefly on the topic
of Carbon Enhanced Metal Poor stars, which is covered
by the review of \cite[Aoki (2009)]{Aoki09} in this 
Symposium. Finally I will largely ignore the abundant literature
on lithium, which should be discussed in  this
volume by the contributions of 
\cite[M\'elendez et al. (2009)]{melendez09},
\cite[Sbordone et al. (2009)]{Sbordone09} and
\cite[Steffen et al. (2009)]{Steffen09}.
I will try to concentrate on the observations, 
without trying to review their theoretical interpretation.

\section{The lowest metallicity stars}

In the search of a ``zero metal'' star, many
Extremely Metal Poor (EMP) stars have been discovered
thanks to the  exploitation of the objective-prism
surveys HK (\cite[Beers et al. 1985]{Beers85},
\cite[Beers et al. 1992]{Beers92}, \cite[Beers 1999]{Beers99}),
Hamburg-ESO
(\cite[Christlieb  2003]{norbert03},\cite[Christlieb et al. 2008]{norbert08})
and, more recently the Sloan Digital Sky Survey (\cite[York et al. 2000]{york}).
Both from the observational and from the theoretical point of view it is important
to establish if there is a threshold in metallicity, below which
no low-mass stars exist. For this reason we would like to know
what is the lowest metallicity found among stars.
There are different answers, depending on how you define
metallicity. The element whose abundance is
most easily measured is iron, so that many people
define metallicity as the iron abundance, or in
spectroscopic notation, [Fe/H]\footnote{[X/Y] = log(X/Y)-log(X/Y)$_\odot$}.
The scientific community became extremely excited by the discovery
of the Hyper Metal Poor stars (HMP, according
to the nomenclature proposed by  \cite[Beers \& Christlieb 2005]{BC05}),
with [Fe/H] of the order of --5, or lower.
The class contains up to now only three
stars, all extracted
from the Hamburg-ESO Survey: HE 0107-5240 ([Fe/H]=--5.3,
\cite[Christlieb et al. 2002]{Christlieb})
HE 1327-2326 ([Fe/H]=--5.4, \cite[Frebel et al. 2005]{Frebel})
and HE 0557-4840 ([Fe/H]=--4.8, \cite[Norris et al. 2007]{Norris}).
A single element would be a fair tracer of the
global metallicity if the element-to-element abundance 
ratios were Universal, but they are not.
The three above stars are characterised by a large overabundance
of C, N and O (see \cite[Collet et al. 2006]{collet06}  for an
analysis of  HE 0107-5240  and HE 1327-2326, based on hydrodynamical
models). This peculiar chemical composition implies that
their metallicity $Z$, the mass fraction of elements
heavier than He, is comparable to that of Globular Clusters
and Halo stars with [Fe/H]$\sim -2.0$, for this reason
I think that the nomenclature proposed 
by  \cite[Beers \& Christlieb (2005)]{BC05} is somewhat
misleading and I suggest
that Hyper Fe Poor stars (HFeP) would be preferable\footnote{Hyper Iron Poor
(HIP) could be confused with Hipparcos numbers.}.
Beyond the purely semantic issue there is obviously the more
fundamental question of the age of these and other EMP stars.
In a naive approach to chemical evolution one expects a well
defined age-metallicity relation and, if so, is a star of 
very low Fe, more ``pristine'' than a star of very low $Z$ ?
The evidence, both in our and external galaxies is that in fact
chemical evolution can be very complex and a simple
age-metallicity relation may not exist. In my view there is 
no compelling evidence that the HFeP stars are more pristine
than other EMP stars and, in fact, all possibilities are open:
they could be older, coeval or younger and they may, indeed,
show a spread in ages. Precise distances from the GAIA mission
(\cite[Perryman et al. 2001 ]{GAIA}) 
will certainly shed new light on this issue.
I would also like to mention the intriguing
evidence shown by \cite[Venn \& Lambert(2008)]{VL},
that the abundance pattern in the HFeP stars is
similar to what observed in dust-forming stars, 
such as post-AGB stars. Whether the HFeP stars
are indeed dusty objects or not can be tested directly
by measuring the abundance of the volatile element S,
and efforts are in progress in this direction.

If we now turn our attention to the extremely low $Z$
stars, the situation is clear, the record holder is 
CD $-38^\circ 245$  discovered by 
\cite[Bessell \& Norris(1984)]{BN84},
with [Fe/H]=--4.2 (\cite[Cayrel et al. 2004]{Cayrel04}),
no measurement of C, N or O, but strong enhancements 
can be excluded, thus 
a value of $Z$ which is of the order of 10$^{-4}$ the solar
value.
There is a handful of giant and sub-giant stars which have
a comparable metallicity: BS 16467-062, CS 22172-002, CS 22885-096
(\cite[Cayrel et al. 2004]{Cayrel04}), BS 16076-006 (\cite[Bonifacio et al. 2007]{B07}),
CS 30336-049 (\cite[Lai et al. 2008]{Lai08})  and 
HE 1424-0241 (\cite[Cohen et al. 2007]{Cohen07}).
The latter star has a markedly different chemical
composition with respect to the others, showing a very low
silicon abundance (1/10 of the iron), but a ``normal'' Mg
abundance. Should oxygen be under-abundant like Si, this would
be the most metal-poor object known.

\section{High resolution surveys}

Several groups have began an homogeneous chemical analysis 
of large numbers of EMP stars, based on data
collected with 8m class telescopes. In this context ``large''
means of the order of a few tens.
The ``First Stars'' group, led by R. Cayrel
has published detailed abundances for giant
(\cite[Cayrel et al. 2004]{Cayrel04})
and dwarf
(\cite[Bonifacio et al. 2009a]{B09})
 stars based on spectra collected with 
UVES at the VLT. The ``0Z project'',
led by J. Cohen, relied on spectra obtained
with HIRES at Keck (\cite[Cohen et al. 2004]{Cohen04}, 
\cite[Cohen et al. 2008]{Cohen08}).
Final the group led by D. Lai has made use of both
ESI (\cite[Lai et al. 2004]{Lai04}) and HIRES
(\cite[Lai et al. 2008]{Lai08}) at Keck.
The good news is that the results of these three
groups agree very well, for the stars in common.
The comparison of the measured equivalent widths is always very good, 
in spite of the differences in observational data
and technique for measurement.
The abundances can differ by up to a factor of two, 
however the differences are well understood in terms
of different atmospheric parameters (obtained with different
methods), different model atmospheres employed, 
different lines selected.
All three groups have published full details of their
analysis, thus making it possible (and perhaps desirable)
a homogeneous analysis of all the available data.
It should be however pointed out that, even without
such an homogenisation, the picture provided by the
abundance ratios measured by each group is highly
consistent.

A special place is held by the HERES survey
(\cite[Christlieb et al. 2004]{Christlieb04},
\cite[Barklem et al. 2005]{Bark05}). By means
of a ``snapshot'' strategy, limited spectral coverage
and medium S/N ratios, it provided detailed abundances
for hundreds of stars.
The chemical information is not as complete or as accurate
as that afforded by the high S/N studies, but the large numbers
involved are indeed highly valuable.
The general picture emerging from the abundance ratios
of the HERES survey is consistent with that coming from
the high S/N studies.

The CASH project (\cite[Frebel et al. 2008a]{Frebel08}) 
is under way at the Hoberly-Eberly telescope and has so far
published the first paper of the series 
(\cite[Roederer et al. 2008]{R08}), but
see also \cite[Roederer et al. (2009)]{R09} in this volume. It is expected
to produce highly interesting results in the next few years.

In the course of these surveys of EMP stars it is only
natural to note some extreme objects, whose chemical composition
departs from that of the vast majority of others,
at the same metallicity. For most of these objects
we do not have a clear idea of the cause for these
peculiar chemical composition.
I already mentioned  HE 1424-0241 and its extraordinarily low
Si abundance. 
Perhaps related to this is SDSS J234723.64+010833.4
(\cite[Lai et al. 2009]{Lai09}) underhanced in Mg
([Mg/Fe]=--0.1) and overenhanced in Ca ([Ca/Fe]=+1.1)
At the opposite side
there is BS 16934-002 (\cite[Aoki et al. 2007]{Aoki07}),
with [Fe/H]=--2.7 and an extreme enhancement
of $\alpha$ elements ([Mg/Fe]=+1.2, [O/Fe]=+1.1)
The giant HK II 17435-00532 (\cite[Roederer et al. 2008]{R08}),
shows an extraordinarily high lithium abundance (A(Li)=2.1) 
and is enhanced in neutron capture elements.
It certainly came as a surprise to me to learn that
the subgiant BD$+44^\circ 493$ (V=9.1) has a metallicity
as low as [Fe/H]=--3.7 (\cite[Ito et al. 2009a]{Ito09}).
The reason why this star has been for so long overlooked
is that it is a CEMP star, thus having a metal-rich appearance
at low resolution. Its brightness allowed to attempt
the measurement of Be. No Be was detected, as 
expected from the linear decrease of Be with metallicity.
The fact that the star shows a measurable Li abundance
(A(Li)=1.04) allows to discard Be destruction      
in the star itself(\cite[Ito et al. 2009b]{Ito265}).
Finally I would like to mention the possible paradox
posed by star CS 30322-023 (\cite[Masseron et al. 2006]{Masseron}),
whose extremely high luminosity (log g $\le -0.3$) qualifies it
as a TP-AGB star. The abundance pattern of this star suggests
an intermediate mass of 2$M_\odot$ or larger.
However, its distance (about 50 kpc) implies it belongs
to the outer Halo, where no recent star formation has occurred.

\section{Highlights of research on EMP stars}

In a somewhat arbitrary manner I want to mention
here some of the results which I think are most exciting.
I will start with Be, this element, which is a pure
product of cosmic ray spallation shows a linear decrease
with metallicity. This has now been confirmed down to
the very lowest metallicities, with no
hint of a ``Be plateau'', by the works of
\cite[Rich \& Boesgaard(2009)]{Rich} and
\cite[Tan et al.(2009)]{Tan}.
On the other hand the large survey
conducted by 
\cite[Smiljanic et al. (2009a)]{rodolfo}
allowed to definitely establish the value
of Be as a chronometer (see also
\cite[Smiljanic et al. 2009b]{rodolfo265}  in this volume). 

For the understanding of the Galactic chemical 
evolution the knowledge of isotopic ratios,
besides that of abundances, provides important
insight. The isotopic ratios
of Li are covered in this volume by 
\cite[Steffen et al. (2009)]{Steffen09} and those
on neutron capture elements by
\cite[Sneden et al. (2009)]{Sneden09}.
I would like here to cite the important progress
which has been made on the
measurement of Mg isotopic ratios
(\cite[Yong et al. 2003]{Yong03}, 
\cite[Yong et al. 2004]{Yong04}, 
\cite[Yong et al. 2006]{Yong06}, 
\cite[Mel{\'e}ndez \& Cohen 2009]{MC09}), which
provide direct evidence of the onset of the contribution
of AGB stars to the chemical evolution.
Such measurements are extremely difficult
and further effort in this
direction is strongly encouraged.

Binary stars always provide us some constraint
on the masses of the components, thus their study
is strongly encouraged. 
They often provide us some puzzles, like
the EMP system   CS 22876-32  ([Fe/H]=--3.6)
for which 
\cite[Gonz{\'a}lez Hern{\'a}ndez et al. (2008)]{jonay08}
have been able to determine the Li abundance
in both components and, surprisingly, the abundance
differs by 0.4\,dex, although the effective temperature of both 
components is too high to expect lithium to be
depleted by convection. 
Another puzzle comes from 
the system CS 22964-161 
(\cite[Thompson et al. 2008]{T08}) in which
both components show a high enhancement in carbon
and s-process elements, as expected if mass-transfer
from an AGB companion had occurred. The puzzle
is that the system is double lined and both stars
appear to be on the Main Sequence. This points to the fact
that this was once a triple system and the most massive
star, after its AGB phase, has in fact been lost.
In this context it is interesting to note
that a quadruple metal-poor system has recently 
been discovered.   
\cite[Rastegaev (2009)]{2009AstL...35..466R} has shown
that G89-14 ([Fe/H]=--1.9) is indeed a highly hierarchical
quadruple system. So perhaps the existence of
a triple system is not so uncommon. 
There is a further anomaly of 
 CS 22964-161, its lithium abundance is A(Li)=2.2,
while we would expect a low value, after the transfer
from an AGB companion, which enhanced the C abundance.
However, this is a feature which is shared
by other CEMP stars, for example 
SDSSJ1036+1212 (\cite[Behara et al. 2009b]{Behara265}
in this volume).

Another extremely exciting finding is that we are now
beginning to find the EMP stars in Local Group galaxies.
The first one found was Draco 119 
(\cite[Shetrone et al. 2001]{Shetrone}),
and the second was found in the Sgr dSph
(\cite[Bonifacio et al. 2006]{bonifacio06}),
however for some time these were considered
the exceptions.
Especially after the DART
collaboration announced a clear lack
of EMP stars in the LG (\cite[Helmi et al. 2006]{helmi})
it was widely felt that these stars
were a peculiarity of the Milky Way.
The situation has now largely changed, 
in the first place the DART collaboration 
revised the metal-poor end of their
calibration of the Ca{\sc ii} IR triplet
(see \cite[Hill 2009]{hill}, these proceedings),
in the second place a number of new EMP stars
has been discovered in LG galaxies.
\cite[Cohen \& Huang(2009)]{CH09} discovered
a second EMP star in Draco, \cite[Frebel et al.(2009)]{Frebel09}
discovered two EMP stars in UMa II and one in 
Coma Berenices, \cite[Norris et al.(2008)]{Norris08} discovered
eight stars with [Fe/H]$\sim -3$ in Bootes I 
and  one with [Fe/H]$\sim -3.5$.
On the other hand Sextans does not show
any stars below [Fe/H]$=-3$, although
\cite[Aoki et al.(2009)]{Aokisextans} found six below --2.5.
The conclusion is that EMP stars are to be found everywhere
and their detailed abundances will tell us something
on the first stars in their host galaxies.

\section{Deviations from LTE}

The bulk of the chemical abundances published to date
assume Local Thermodynamic Equilibrium (LTE)
in the line formation computations.
We know that this is an idealised assumption and 
there is a very active research on relaxing it.

The odd elements Na and Al show sizeable NLTE
effects and all abundances based on LTE analysis
should be discarded (\cite[Gehren et al. 2006]{G06},
\cite[Andrievsky et al. 2007]{A07}, \cite[Andrievsky et al. 2008]{A08}).
Magnesium shows a dwarf/giant discrepancy and should
also be treated in NLTE (\cite[Gehren et al. 2006]{G06}, \cite[Spite et al. 2009]{Spite09}).
The trend of [Mg/Fe] with metallicity is flat in both cases, but higher
in  NLTE ($\sim 0.6$\,dex) than in LTE. In fact when Mg is compute in 
NLTE [O/Mg]$\sim 0$ at all metallicities (\cite[Spite et al. 2009]{Spite09}).
Silicon is also one of the elements which shows a disturbing
dwarf/giant discrepancy (\cite[Bonifacio et al. 2009a]{B09}) and the computations 
of \cite[Shi et al.(2009)]{Shi09} suggest that NLTE is indeed
important for Si in metal-poor stars. 
A thorough NLTE analysis of Si in EMP stars is strongly encouraged.

Carbon is also an element which shows a dwarf/giant
discrepancy (\cite[Bonifacio et al. 2009a]{B09}),
although in this case the discrepancy might have
an astrophysical cause (modification of the abundances
of giant  stars due to mixing) it is more likely that
it is due to an inadequacy in the analysis.
C abundances in metal-poor stars rely mainly on the G-band
and up to now NLTE analysis of CH lines have not been published.
Such an investigation, however is strongly encouraged.

Some abundance ratios
have been discovered early on, to depart significantly
from the solar value in metal-poor stars. For example
\cite[McWilliam et al.(1995)]{McW} found that the [Cr/Fe] and [Mn/Fe]
ratios become increasingly lower for the most metal-poor stars,
while the [Co/Fe] ratios increase.
These findings were consistently confirmed with lower and lower
scatter, by all subsequent investigations and were
generally interpreted as 
features of the Galactic chemical evolution
(see e.g. \cite[Prantzos 2008]{Prantzos} and references
therein).
However it now appears very likely that the trend of
Cr is spurious and due to the neglect of NLTE
effects. As pointed out by \cite[Lai et al.(2008)]{Lai08}
and \cite[Bonifacio et al.(2009a)]{B09}, 
there is a discrepancy between dwarfs and giants,
and if only Cr{\sc ii} lines are used (possible only for giants)
[Cr/Fe] appears to be consistently solar at all metallicities.
Theoretical computations by \cite[Bergemann \& Gehren (2009)]{BG265}
confirm that a NLTE analysis implies a solar [Cr/Fe].
A similar dwarf/giant discrepancy
is present also for Mn (\cite[Bonifacio et al. 2009a]{B09})
and the NLTE computations of \cite[Bergemann \& Gehren(2008)]{BG08}
indeed confirm that the trend is spurious.
Also the [Co/Fe] ratio displays a dwarf/giant discrepancy
(\cite[Bonifacio et al. 2009a]{B09}), however in this
case, the NLTE analysis of 
\cite[Bergemann et al.(2009)]{BG09} implies and even
steeper increase of this ratio with decreasing metallicities.
The NLTE trend for [Cr/Fe] (flat at solar metallicity)
certainly goes in the direction to satisfy chemical
evolution models, and associated stellar yields.
On the contrary the NLTE trends of Mn and Co cannot
be explained by current models.

The situation for copper is unclear.
At very low metallicity the copper abundances must rely
on the strong resonance lines of Mult.\,1  
and the discrepancy of the abundances derived from
these lines and those derived from those
of Mult.\,2 cast serious doubts on the validity
of LTE for either multiplet (\cite[Bonifacio et al. 2009b]{Cu}).

For zinc the situation is puzzling. 
\cite[Bonifacio et al.(2009a)]{B09} found a
disturbing dwarf/giant discrepancy, however
the NLTE computations of 
\cite[Takeda et al.(2005)]{Takeda} imply
small NLTE corrections for Zn and in any case
not significantly different for giants and 
dwarfs. Also
granulation effects are unlikely to be the cause
of this discrepancy. In the case of zinc we are
also facing the problem of small number statistics,
since for very few dwarf stars zinc has been measured.
The issue should be further investigated both from 
the theoretical and observational point of view.

The NLTE effects may be relevant also for 
neutron capture elements.
Up to now results for Ba have been published 
(\cite[Mashonkina et al. 2008]{Mashonkina}, \cite[Andrievsky et al. 2009]{A09})
and in this case the large star-to-star scatter at any metallicity
is confirmed by the NLTE analysis, pointing to 
a poor mixing of these elements in the early Galaxy.

\begin{figure}
\begin{center}
\resizebox{\hsize}{!}{\includegraphics[clip=true]{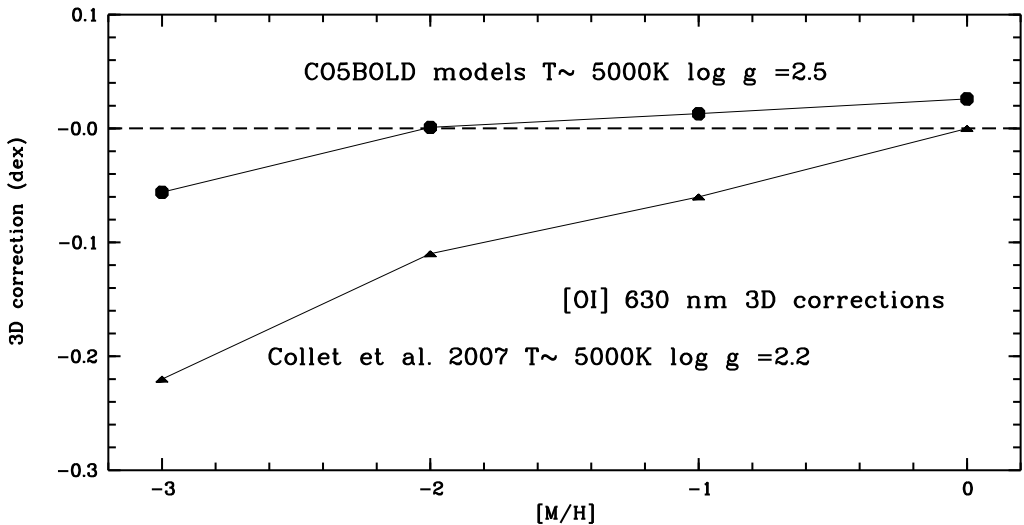}}
\caption{3D corrections for the [OI] 630\,nm line, for giant stars
computed from \cobold\ models, compared with those computed
by \cite[Collet et al. (2007)]{collet}, from models
computed with the Stein\& Nordlund code (\cite[Stein \& Nordlund 1998]{SN98}). 
The difference is non-negligible, and it is likely rooted
in the different microphysics of the two codes, but also 
in the different models used as 1D reference to compute the 3D corrections.
}
   \label{fig1}
\end{center}
\end{figure}

\section{Granulation effects}

Besides LTE, the most important simplifying
assumption made in the analysis of stellar spectra
is that of a static atmosphere. Thus the majority of analysis
rely on 1D hydrostatic model atmospheres.
In the last 10 years a considerable advance
has come through the use of three dimensional
hydrodynamical simulations of stellar atmospheres (3D models
for short).
All the analysis published so far rely on simulations
computed either with the 
code of \cite[Stein \& Nordlund(1998)]{SN98} or 
with the \cobold\ code 
(\cite[{{Freytag} {et~al. }2002}]{Freytag2002AN....323..213F},
\cite[{{Freytag} {et~al. }2003}]{Freytag2003CO5BOLD-Manual},
\cite[{{Wedemeyer} {et~al. }2004}]{Wedemeyer2004A&A...414.1121W}).
Such models are more physically motivated than 1D models,
although there is still considerable work to validate them
and bring them at the level of maturity of current 1D models.
The treatment of opacity in such models is based on 
an opacity binning scheme (\cite[Nordlund 1982]{Nordlund82},
\cite[Ludwig 1992]{Ludwig92},
\cite[Ludwig, Jordan, \& Steffen 1994]{Ludwig+al94}),
however the optimal number of bins to employ and their
definition is still a matter of investigation.
\cite[Behara et al.(2009a)]{Behara} found significant differences
in the temperature structure of the outer layers for models
computed using six or twelve opacity bins.

To illustrate some of the problems I computed the ``3D correction'',
as defined by \cite[Caffau \& Ludwig(2007)]{zolfito}, for 
for the [OI] 630\,nm line, from 
four models
of giant star extracted from the CIFIST grid of 3D models (\cite[Ludwig et al. 2009]{L09}).  
In Fig.~\ref{fig1} I compare these corrections with those 
published by \cite[Collet et al. (2007)]{collet}.
The difference is small, but non-negligible in the context of Galactic
chemical evolution. My computations suggest that the 1D-based oxygen 
abundances of \cite[Cayrel et al.(2004)]{Cayrel04} require
no correction for granulation effects, 
while the computations of  \cite[Collet et al. (2007)]{collet}
imply a downward revision by 0.2\,dex or perhaps larger.
At the time of writing I am unable to say which of the
two computations is right (if any !).
I can however point out two differences which are likely
to be relevant: {\em i) }the models of the CIFIST grid are 
computed using six opacity bins, while those
of \cite[Collet et al. (2007)]{collet} use four opacity bins;
{\em ii)} my corrections are computed using as reference
1D model an LHD model (see e.g. \cite[Caffau et al. 2008]{oxy}),
which employs the same microphysics of \cobold , while
\cite[Collet et al. (2007)]{collet} use a MARCS model 
(\cite[Gustafsson et al. 2008]{marcs}, and references therein).
The role of these differences still needs to be explored.

It should be clear that the choice of using
to hydrodynamical models forces us to make some
simplifications which are not done in 1D hydrostatic
models. The most obvious one is the role of scattering.
While this is properly treated in existing 1D model
atmosphere and line formation codes, it is treated
as true absorption in all 3D codes. 
It remains to be investigated if this 
approximation is acceptable or not. 

In the meanwhile it is exciting to note a 
strong effort in a systematic application 
of  3D models to abundance analysis
(\cite[Collet et al. 2006]{collet06},
\cite[Collet et al. 2007]{collet},
\cite[Cayrel et al. 2007]{Cayrel07},
\cite[Gonz{\'a}lez Hern{\'a}ndez et al. 2008]{jonay08},
\cite[Frebel et al. 2008b]{Frebel08b},
\cite[Collet et al. 2009]{collet09}, 
\cite[Bonifacio et al. 2009a]{B09} , 
\cite[Gonz{\'a}lez Hern{\'a}ndez et al. 2009]{jonay09}
)

All the aspects of spectroscopic analysis need to be 
explored and revised in the light of hydrodynamical models.
One noticeable example are the Balmer lines and their
role in temperature diagnostic (\cite[Ludwig et al. 2009]{balmer}).
One of the things that we still lack from 3D models
are extensive grids of theoretical fluxes and colours,
although efforts in this direction are underway
(\cite[Ku\v{c}inskas et al. 2009]{K09},\cite[Casagrande 2009]{Casagrande})

The future looks very bright and busy, for the vast number
of tasks to be accomplished. I hope the community
will continue with the enthusiasm shown in the last decade.

\begin{acknowledgments}
I am grateful to all the colleagues
who have helped in preparing this review and
in particular to 
Hans G\"unter Ludwig and Elisabetta Caffau,
who also provided me the hydrodynamical models
used for the computations, 
Monique Spite, Fran\c cois Spite, Roger Cayrel
and Chris Sneden.
A special thanks to Katia Cunha, for her patience
in managing my manuscript.
I acknowledge financial
support from EU contract MEXT-CT-2004-014265 (CIFIST).
\end{acknowledgments}


\begin{thebibliography}{}

\bibitem[Andrievsky et al.(2007)]{A07} Andrievsky, S.~M., Spite, M., 
Korotin, S.~A., Spite, F., Bonifacio, P., Cayrel, R., Hill, V., \& Fran{\c c}ois, P.\ 2007, \aap, 464, 1081 



\bibitem[Andrievsky et al.(2008)]{A08} Andrievsky, S.~M., Spite, M., 
Korotin, S.~A., Spite, F., Bonifacio, P., Cayrel, R., Hill, V., \& Fran{\c c}ois, P.\ 2008, \aap, 481, 481 

\bibitem[Andrievsky et al.(2009)]{A09} Andrievsky, S.~M., Spite, M., 
Korotin, S.~A., Spite, F., Fran{\c c}ois, P., Bonifacio, P., Cayrel, R., \& Hill, V.\ 2009, \aap, 494, 1083 



\bibitem[Aoki (2009)]{Aoki09} Aoki, W.\ 2009 \textit{IAU Symp.} 265, p. 111

\bibitem[Aoki et al.(2007)]{Aoki07} Aoki, W., et al.\ 2007, 
\apj, 660, 747 

\bibitem[Aoki et al.(2009)]{Aokisextans} Aoki, W., et al.\ 2009, \aap, 502, 569 


\bibitem[Barklem et al.(2005)]{Bark05} Barklem, P.~S., et al.\ 2005, \aap, 439, 129 




\bibitem[Beers (1999)]{beers99} Beers, T.~C.\ 1999, \apss, 265, 547 



\bibitem[Beers \& Christlieb(2005)]{BC05} Beers, T.~C., \& Christlieb, N.\ 2005, \araa, 43, 531 


\bibitem[Beers et al. (1985)]{beers85} Beers, T.~C., Preston, 
G.~W., \& Shectman, S.~A.\ 1985, \aj, 90, 2089 

\bibitem[Beers et al. (1992)]{beers92} Beers, T.~C., Preston, 
G.~W., \& Shectman, S.~A.\ 1992, \aj, 103, 1987 


\bibitem[Behara et al.(2009a)]{Behara} Behara, N.~T., Ludwig, 
H.~-G., Bonifacio, P., Sbordone, L., Gonz\'alez Hern\'andez, J.~I., 
\& Caffau, E.\ 2009a, \memsai, 80, 732  

\bibitem[Behara et al.(2009b)]{Behara265} Behara, N.~T., 
Bonifacio, P., Ludwig, H.~-., Sbordone, L., Gonz\'alez Hern\'andez, J.~I., 
\& Caffau, E.\ 2009b,  \textit{IAU Symp.} 265, p. 122


\bibitem[Bergemann \& Gehren(2008)]{BG08} Bergemann, M., \& Gehren, T.\ 2008, \aap, 492, 823 

\bibitem[Bergemann \& Gehren (2009)]{BG265} Bergemann, M., \&
Gehren, T.\ 2009,  \textit{IAU Symp.} 265, p. 348 

\bibitem[Bergemann et al.(2009)]{BG09} Bergemann, M., 
Pickering, J.~C., \& Gehren, T.\ 2009, arXiv:0909.2178 


\bibitem[Bessell \& Norris(1984)]{BN84} Bessell, M.~S., \& Norris, J.\ 1984, \apj, 285, 622

\bibitem[Bonifacio et al.(2006)]{bonifacio06} Bonifacio, P., et 
al.\ 2006, \textit{Chemical Abundances and Mixing in Stars in the Milky Way and its 
Satellites, ESO ASTROPHYSICS SYMPOSIA.}~ISBN 
978-3-540-34135-2.~Springer-Verlag, 2006, p.~232

\bibitem[Bonifacio(2007)]{Bonifacio07} Bonifacio, P.\ 2007, \textit{EAS Publications Series}, 24, 251 

\bibitem[Bonifacio et al.(2007)]{B07} Bonifacio, P., et al.\ 2007, \aap, 462, 851

\bibitem[Bonifacio et al.(2009a)]{B09} Bonifacio, P., et al.\ 2009a, \aap, 501, 519 

\bibitem[Bonifacio et al.(2009b)]{Cu} Bonifacio, P., 
Caffau, E., \& Ludwig, H.~-.\ 2009b, \memsai 80, 736

\bibitem[Caffau \& Ludwig(2007)]{zolfito} Caffau, E., \& Ludwig, H.-G.\ 2007, \aap, 467, L11 

\bibitem[Caffau et al.(2008)]{oxy} Caffau, E., Ludwig, H.-G., 
Steffen, M., Ayres, T.~R., Bonifacio, P., Cayrel, R., Freytag, B., \& Plez, B.\ 2008, \aap, 488, 1031 

\bibitem[Casagrande (2009)]{Casagrande} Casagrande, L.\ 2009, \memsai, 80, 724

\bibitem[Cayrel et al.(2004)]{Cayrel04} Cayrel, R., et al.\ 2004, \aap, 416, 1117 

\bibitem[Cayrel et al.(2007)]{Cayrel07} Cayrel, R., et al.\ 2007, \aap, 473, L37 

\bibitem[Christlieb(2003)]{norbert03} Christlieb, N.\ 2003, 
\textit{Reviews in Modern Astronomy}, 16, 191

\bibitem[Christlieb et al.(2002)]{Christlieb} Christlieb, N., et 
al.\ 2002, \nat, 419, 904  

\bibitem[Christlieb et al.(2004)]{Christlieb04} Christlieb, N., et al.\ 2004, \aap, 428, 1027

\bibitem[Christlieb et al.(2008)]{norbert08} Christlieb, N., 
Sch{\"o}rck, T., Frebel, A., Beers, T.~C., Wisotzki, L., \& Reimers, D.\ 2008, \aap, 484, 721 



\bibitem[Cohen \& Huang(2009)]{CH09} Cohen, J.~G., \& Huang, W.\ 2009, \apj, 701, 1053 

\bibitem[Cohen et al.(2004)]{Cohen04} Cohen, J.~G., et al.\ 
2004, \apj, 612, 1107 

\bibitem[Cohen et al.(2007)]{Cohen07} Cohen, J.~G., McWilliam, 
A., Christlieb, N., Shectman, S., Thompson, I., Melendez, J., Wisotzki, L., 
\& Reimers, D.\ 2007, \apjl, 659, L161 

\bibitem[Cohen et al.(2008)]{Cohen08} Cohen, J.~G., Christlieb, 
N., McWilliam, A., Shectman, S., Thompson, I., Melendez, J., Wisotzki, L., 
\& Reimers, D.\ 2008, \apj, 672, 320 



\bibitem[Collet et al.(2006)]{collet06} Collet, R., Asplund, M., 
\& Trampedach, R.\ 2006, \apjl, 644, L121 

\bibitem[Collet et al.(2007)]{collet} Collet, R., Asplund, M., \& Trampedach, R.\ 2007, \aap, 469, 687 

\bibitem[Collet et al.(2009)]{collet09} Collet, R., Nordlund, 
\AA., Asplund, M., Hayek, W., \& Trampedach, R.\ 2009, \memsai, 80, 716 




\bibitem[Frebel et al.(2005)]{Frebel} Frebel, A., et al.\ 
2005, \nat, 434, 871 

\bibitem[Frebel et al.(2008a)]{Frebel08} Frebel, A., Allende 
Prieto, C., Roederer, I.~U., Shetrone, M., Rhee, J., Sneden, C., Beers, 
T.~C., \& Cowan, J.~J.\ 2008a, \textit{New Horizons in Astronomy, ASPC}, 393, 203 


\bibitem[Frebel et al.(2008b)]{Frebel08b} Frebel, A., Collet, R., 
Eriksson, K., Christlieb, N., \& Aoki, W.\ 2008b, \apj, 684, 588 

\bibitem[Frebel et al.(2009)]{Frebel09} Frebel, A., Simon, 
J.~D., Geha, M., \& Willman, B.\ 2009, arXiv:0902.2395 


\bibitem[{{Freytag} {et~al.}(2002)}]{Freytag2002AN....323..213F}
{Freytag}, B., {Steffen}, M., \& {Dorch}, B. 2002, \textit{Astronomische Nachrichten},
  323, 213


\bibitem[{{Freytag} {et~al.}(2003)}]{Freytag2003CO5BOLD-Manual}
{Freytag}, B., {Steffen}, M., {Wedemeyer-B{\"o}hm}, S., \& {Ludwig}, H.-G.
  2003, \textit{CO5BOLD User Manual},
  \href{http://www.astro.uu.se/~bf/co5bold_main.html}{http://www.astro.uu.se/~bf/co5bold\_main.html}

\bibitem[Gehren et al.(2006)]{G06} Gehren, T., Shi, J.~R., 
Zhang, H.~W., Zhao, G., \& Korn, A.~J.\ 2006, \aap, 451, 1065 



\bibitem[Gonz{\'a}lez Hern{\'a}ndez et al.(2008)]{jonay08} Gonz{\'a}lez Hern{\'a}ndez, J.~I., et al.\ 2008, \aap, 480, 233 

\bibitem[Gonz{\'a}lez Hern{\'a}ndez et al.(2009)]{jonay09} Gonz{\'a}lez Hern{\'a}ndez, J.~I., et al.\ 2009, \aap, 505, L13 


\bibitem[Gustafsson et al.(2008)]{marcs} Gustafsson, B., Edvardsson, B., 
Eriksson, K., J{\o}rgensen, U.~G., Nordlund, {\AA}., \& Plez, B.\ 2008, \aap, 486, 951 


\bibitem[Helmi et al.(2006)]{helmi} Helmi, A., et al.\ 2006, 
\apjl, 651, L121 

\bibitem[Hill (2009)]{hill} Hill, V.\ 2009, \textit{IAU Symp.} 265, p. 219
 

\bibitem[Ito et al.(2009a)]{Ito09} Ito, H., Aoki, W., Honda, 
S., \& Beers, T.~C.\ 2009a, \apjl, 698, L37 

\bibitem[Ito et al.(2009b)]{Ito265} Ito, H., Aoki, W., Honda, 
S., Beers, T.~C., \& Tominaga, N.\ 2009b,  \textit{IAU Symp.} 265, p. 124

\bibitem[Ku\v{c}inskas et al.(2009)]{K09} Ku\v{c}inskas, A., 
Ludwig, H.~-G., Caffau, E., \& Steffen, M.\ 2009, \memsai, 80, 720 


\bibitem[Lai et al.(2004)]{Lai04} Lai, D.~K., Bolte, M., 
Johnson, J.~A., \& Lucatello, S.\ 2004, \aj, 128, 2402 

\bibitem[Lai et al.(2008)]{Lai08} Lai, D.~K., Bolte, M., 
Johnson, J.~A., Lucatello, S., Heger, A., 
\& Woosley, S.~E.\ 2008, \apj, 681, 1524 

\bibitem[Lai et al.(2009)]{Lai09} Lai, D.~K., Rockosi, C.~M., 
Bolte, M., Johnson, J.~A., Beers, T.~C., Lee, Y.~S., Allende Prieto, C., 
\& Yanny, B.\ 2009, \apjl, 697, L63 


\bibitem[Ludwig (1992)]{Ludwig92}
Ludwig, H.-G. 1992, PhDT, University of Kiel

\bibitem[Ludwig, Jordan,
\& Steffen (1994)]{Ludwig+al94} Ludwig, H.-G., Jordan, S., \& Steffen M.
1994, \aap, 284, 105

\bibitem[Ludwig et al.(2009)]{L09} Ludwig, H.~-., Caffau, 
E., Steffen, M., Freytag, B., Bonifacio, P., 
\& Ku\v{c}inskas, A.\ 2009, \memsai, 80, 708 

\bibitem[Ludwig et al.(2009)]{balmer} Ludwig, H.-G., 
Behara, N.~T., Steffen, M., \& Bonifacio, P.\ 2009, \aap, 502, L1 

\bibitem[Mashonkina et al.(2008)]{Mashonkina} Mashonkina, L., et al.\ 2008, \aap, 478, 529 

\bibitem[Masseron et al.(2006)]{Masseron} Masseron, T., et al.\ 2006, \aap, 455, 1059 


\bibitem[McWilliam et al.(1995)]{McW} McWilliam, A., 
Preston, G.~W., Sneden, C., \& Searle, L.\ 1995, \aj, 109, 2757 

\bibitem[Mel{\'e}ndez \& Cohen(2009)]{MC09} Mel{\'e}ndez, J., \& Cohen, J.~G.\ 2009, \apj, 699, 2017 


\bibitem[M\'elendez et al. (2009)]{melendez09} M\'elendez, J.,  
Casagrande, L., Ramirez, I.\ 2009 \textit{IAU Symp.} 265, p.  71

\bibitem[Nordlund (1982)]{Nordlund82}
  Nordlund, \AA. 1982, \aap, 107, 1

\bibitem[Norris et al.(2007)]{Norris} Norris, J.~E., 
Christlieb, N., Korn, A.~J., Eriksson, K., Bessell, M.~S., Beers, T.~C., 
Wisotzki, L., \& Reimers, D.\ 2007, \apj, 670, 774 

\bibitem[Norris et al.(2008)]{Norris08} Norris, J.~E., Gilmore, 
G., Wyse, R.~F.~G., Wilkinson, M.~I., Belokurov, V., Evans, N.~W., 
\& Zucker, D.~B.\ 2008, \apjl, 689, L113 

\bibitem[Perryman et al.(2001)]{GAIA} Perryman, M.~A.~C., et al.\ 2001, \aap, 369, 339 

\bibitem[Prantzos(2008)]{Prantzos} Prantzos, N.\ 2008, \textit{EAS 
Publications Series}, 32, 311 

\bibitem[Rastegaev(2009)]{2009AstL...35..466R} Rastegaev, D.~A.\ 2009, 
\textit{Astronomy Letters}, 35, 466 

\bibitem[Rich \& Boesgaard(2009)]{Rich} Rich, J.~A., \& Boesgaard, A.~M.\ 2009, \apj, 701, 1519 

\bibitem[Roederer et al.(2008)]{R08} Roederer, I.~U., et 
al.\ 2008, \apj, 679, 1549 

\bibitem[Roederer et al. (2009)]{R09}  Roederer, I.~U.\ 2009 \textit{IAU Symp.} 265, p.  368

\bibitem[Sbordone et al. (2009)]{Sbordone09} Sbordone, L., 
Bonifacio, P.,  Caffau, E., et al. \ 2009 \textit{IAU Symp.} 265, p. 75 

\bibitem[Shetrone et al.(2001)]{Shetrone} Shetrone, M.~D., 
C{\^o}t{\'e}, P., \& Sargent, W.~L.~W.\ 2001, \apj, 548, 592 

\bibitem[Shi et al.(2009)]{Shi09} Shi, J.~R., Gehren, T., Mashonkina, L., \& Zhao, G.\ 2009, \aap, 503, 533 



\bibitem[Smiljanic et al.(2009a)]{rodolfo} Smiljanic, R., Pasquini, L., 
Bonifacio, P., Galli, D., Gratton, R.~G., Randich, S., \& Wolff, B.\ 2009a, \aap, 499, 103 

\bibitem[Smiljanic et al.(2009b)]{rodolfo265} Smiljanic, R., 
Pasquini, L., Bonifacio, P., Galli, D., Barbuy, B., Gratton, R., 
\& Randich, S.\ 2009b,   \textit{IAU Symp.} 265, p.  134

\bibitem[Sneden et al. (2009)]{Sneden09} Sneden, C., Cowan, J.J. \& Gallino, R.\ 2009 \textit{IAU Symp.} 265, p. 46 

\bibitem[Sneden et al.(2008)]{Sneden08} Sneden, C., Cowan, J.~J., \& Gallino, R.\ 2008, \araa, 46, 241 

\bibitem[Spite et al. (2009)]{Spite09} Spite, M., Spite, F., Bonifacio, P., Andrievsky, S., Cayrel R., 
Fran\c cois, P., Korotin, S.\ 2009 \textit{IAU Symp.} 265, p. 380 

\bibitem[Steffen et al. (2009)]{Steffen09} Steffen, M., Cayrel, R.,
Bonifacio, P., Ludwig, H.-G., Caffau, E.\ 2009 \textit{IAU Symp. 265}, p. 23 


\bibitem[Stein \& Nordlund(1998)]{SN98} Stein, R.~F., \& Nordlund, A.\ 1998, \apj, 499, 914 

\bibitem[Takeda et al.(2005)]{Takeda} Takeda, Y., Hashimoto, 
O., Taguchi, H., Yoshioka, K., Takada-Hidai, M., Saito, Y., 
\& Honda, S.\ 2005, \pasj, 57, 751 

\bibitem[Tan et al.(2009)]{Tan} Tan, K.~F., Shi, J.~R., 
\& Zhao, G.\ 2009, \mnras, 392, 205 

\bibitem[Thompson et al.(2008)]{T08} Thompson, I.~B., et 
al.\ 2008, \apj, 677, 556 

\bibitem[Venn \& Lambert(2008)]{VL} Venn, K.~A., \& Lambert, D.~L.\ 2008, \apj, 677, 572 

\bibitem[{{Wedemeyer} {et~al.}(2004)}]{Wedemeyer2004A&A...414.1121W}
{Wedemeyer}, S., {Freytag}, B., {Steffen}, M., {Ludwig}, H.-G., \& {Holweger},
  H. 2004, \aap, 414, 1121


\bibitem[Yong et al.(2003)]{Yong03} Yong, D., Lambert, D.~L., 
\& Ivans, I.~I.\ 2003, \apj, 599, 1357 

\bibitem[Yong et al.(2004)]{Yong04} Yong, D., Lambert, D.~L., 
Allende Prieto, C., \& Paulson, D.~B.\ 2004, \apj, 603, 697 



\bibitem[Yong et al.(2006)]{Yong06} Yong, D., Aoki, W., 
\& Lambert, D.~L.\ 2006, \apj, 638, 1018 

\bibitem[York et al.(2000)]{york} York, D.~G., et al.\ 2000, 
\aj, 120, 1579 



\end{thebibliography}
\end{document}